# Few layer graphene from mechanical exfoliation of graphite based materials. Structure-dependent characteristics.

Azhar A. Pirzado [1,2], François Le Normand [2], Thierry Romero,[1] Sandra Paszkiewicz[3], Vasiliki Papaefthimiou[1], Dris Ihiawakrim[4], and Izabela Janowska[1*]

[1] Institut de Chimie et Procédés pour l'Énergie, l'Environnement et la Santé (ICPEES), CNRS UMR 7515- University of Strasbourg, 25 rue Becquerel 67087 Strasbourg, France; pirzado@etu.unistra.fr, thierry.romero@unistra.fr, papaefthymiou@unistra.fr, janowskai@unistra.fr
[2] Laboratoire des sciences de l'Ingénieur, de l'Informatique et de l'Imagerie (ICube), UMR 7357, CNRS – University of Strasbourg, 67400 Strasbourg, France ; francois.le-normand@unistra.fr
[3] West Pomeranian University of Technology, Institute of Material Science and Engineering Piastow Av. 19, 70310 Szczecin, Poland; spaszkiewicz@zut.edu.pl,
[4] Institut de Physique et Chimie des Matériaux de Strasbourg (IPCMS), CNRS UMR 7504-University of Strasbourg, France ; drisihi@ipcms.unistra.fr

* Correspondence: : *janowskai@unistra.fr*; Tel.: +33 (0)36-885-2633

**Abstract:** We present a high scale method to produce few layer graphene (FLG) based on the mechanical exfoliation of graphite and compare the obtained FLG with the one reported earlier arising from pencil lead ablation [1]. Several things are modified and improved in the new approach. The purification and the ablation set-up are simplified, and the morphology of the FLG is modified and improved in view of some applications. The morphology dependent properties of FLGs, lead-FLG and graphite-FLG, are investigated as conductive layers and in nanocomposites. Newly obtained FLG has higher aspect ratio (high lateral size vs. thickness/higer 2D aspect) which is reflected by an enhanced transparency-conductivity features of the layer (film) and an elongation at break behavior in the polymer composites. On the contrary, the nanocomposite containing lead-FLG shows for instance excellent gas barrier properties due to the multi-step structure of lead-FLG flakes. Such structure exhibits less 2D and more 3D character, which can be highly suitable for applications where the presence of active/reactive edges is beneficial, e.g. in catalysis or supercapacitors' electrodes. Nuclear reaction analysis is employed to investigate the morphology of graphite-FLG film.

## 1. Introduction

A significant number of methods have been developed for the production of graphene and/or few layer graphene (FL)G since graphene has been mechanically isolated from graphite by a scotch tape for the first time in 2004 [2]. The ideal method should fit the envisaged applications in term of cost, efficiency, scalability but also the morphology-properties of the obtained (FL)G product. High expectations from industry concerning the applications of graphene and graphene based materials induce also a permanent search for efficient methods that additionally could answer the sustainable development requirements. The bottom-up methods such as CVD or epitaxial growth over SiC produce high quality graphene films dedicated to more sophisticated needs in electronics, optoelectronics, spintronics and related fields [3, 4], while their limited yields and scalability exclude them from many other applications. The fields that require high-yield "bulk graphene"/(FL)G powder/platelets are for instance high performance composites, conductive inks, coatings or catalysis. They focus on top-down approaches dealing with the exfoliation of graphite based materials, where the van der Waals forces between sheets in graphite need to be defeated. The most common top-down method is the ultrasonication-assisted exfoliation in liquids, either in an organic solvent with the appropriate surface tension [5] or in aqua solution in a presence of surfactant. Additionally, a large number of works reporting the production of reduced graphene oxide focus mostly on the restoration of the continuous conjugated -C=C- lattice lost during the highly oxidative exfoliation of graphite into the graphite oxide [6]. The recent prominent exfoliation in water rather involves the use of bio-surfactants [7, 8], where high exfoliation yield and highly concentrated colloids can be reached [8]. Apart from exfoliation improvement by chemistry, the mechanically applied shear forces and the way they are applied play a significant role, impacting the efficiency and structure of the resulting (FL)G. A significant progress of the (FL)graphene production was observed, for instance, when a kitchen blender [9], microfluidization [10] or mixing-assisted ultrasonication were applied [8].

The scarce top-down methods concern the mechanical exfoliation of graphite, including the scotch tape approach, and usually either proceed with low yield or lead to (FL)G with a highly disturbed morphology as it is the case for ball-milled graphite [2, 11, 12]. One of the highest yield method reported by us a few years ago was the mechanical ablation of a pencil lead on a rough glass surface [1]. The ablation was assisted by the simultaneous ultrasonication of the glass surface in ethanol (water) aiming at the detachment of the ablated (exfoliated products).

Herein, we show several beneficial modifications of this mechanical ablation method that were conducted by replacing the pencil lead by graphite discs with an adaption of the processing and set-up.

## 2. Materials and Methods

Lead-FLG has been obtained according to the work reported previously [1].

Graphite-FLG has been obtained through the mechanical ablation of pure graphite disc (provided by Nanocyl) on a rough glass surface. The set-up consisted of modified polishing-like apparatus, in which the horizontal glass disc as ablation substrate turned with a controlled speed, while the graphite disc hanged loosely over glass surface kept by a suction cup system from the top. A stream of ethanol was systematically used to remove the material ablated and attached to the glass surface. The product of the ethanol suspension was next submitted to a sedimentation process for 4 h. The supernatant containing FLG was then separated with an overall yield of around 60 % and dried under vacuum evaporation. Eventually a final drying in a standard oven at 100 °C for 2 h was applied.

FLG layers/films have been prepared from ethanol suspensions with concentrations of 0.5 mg/mL that were sprayed on the preheated and continuously heated quartz plates at 120°C with Airgun (Hi-line Iwata). Air was used as a carrier gas with gun inlet pressure of 1.5 bar.

Thermal annealing of the films was performed at 900°C for 2h under Ar flow.

Transmission electron microscopy (TEM) was carried out on a Topcon 002B - UHR microscope working with an accelerated voltage of 200 kV and a point - to -point resolution of 0.17 nm. The sample was dispersed in ethanol and a drop of the suspension was deposited onto a carbon coated copper grid for analysis. In the case of composite, a slice of composite was cut prior to analysis.

Scanning electron microscopy (SEM) analysis was carried out on JEOL 6700-FEG microscope.

TGA analyses were carried out on TA instrument SDT Q600 under Air, the rate of heating was fixed at 5°/min.

Raman spectroscopy was performed using Horiba Scientific Labram Aramis Raman Spectrometer (JobinYvon technolgy) with the following conditions: laser wavelength of 532.15 nm, D2 filter (1% power) and spectrum in regions from 1250 to 1650 and from 2600 to 2800 cm$^{-1}$, with integration time of 100s for each phase.

The XPS measurements were performed in an ultrahigh vacuum (UHV) setup equipped with a VSW ClassWA hemispherical electron analyzer with a multi-channeltron detector. A monochromated AlK$\alpha$ X-ray source (1486.6 eV; anode operating at 240 W) was used as incident radiation. The base chamber pressure was $1 \times 10^{-9}$ mbar. High- resolution spectra were recorded in constant pass energy mode (100 and 20 eV, respectively). Prior to individual elemental scans, a survey scan was taken for all the samples to detect all of the present elements.

UPS measurements. During the work function measurements by UPS ($h\nu$=21.2 eV), a bias of 15.31 V was applied to the samples, in order to avoid interference of the spectrometer threshold in the UP spectra. The work function of the surface is determined from the UP spectra by subtracting their width (i.e., the energy difference between the Fermi level and the high binding energy cutoff) from the He I excitation energy (21.2 eV).

Transmission spectra (UV-Vis) were obtained on VARIAN Cary 100 Scan UV-Visible spectrometer having a deuterium arc and tungsten halogen lamps.

The thickness of the graphite-FLG was measured by a Veeco DEKTAK 150 profilometer.

Four point probes measurements (FPPs) has been performed in order to achieve sheet resistance ($R_s$) of the lead-FLG and graphite-FLG films, using Keithley 220 programmable current source coupled with a Hewlett-Packard 34401A multimeter. For some films, a pair of electrodes (Au/Cr) has been deposited. It was however noticed by the systemically performed measurements that the electrodes are not necessary to attain good contact with FLG films and the measurable values were comparable with electrodes-free films.

The average thickness of the films was measured by Dektac profilometer (and confirmed by atomic force microscopy).

Nuclear reaction analysis (NRA) analysis. The ion beam analysis was carried out on 4MV Van De Graff accelerator facility, using deuterium (2H) as an incident ion with an energy of 900 keV and a scattering angle of 150°. The conversion from channel (C) to energy (E) in keV (arbitrary number slots converted to corresponding energy values, E = 7.04 × C-85.00) is done by using the standard oxygen and carbon nuclear reactions [$O^{16}(d,p_0)O^{17}$], [$O^{16}(d,p_1)O^{17}$] and [$C^{12}(d,p)C^{13}$] . The bare SiO$_2$ and SiC samples were used for the calibration of the spectra. The spectra were calibrated by using SiO$_2$ and SiC samples for oxygen and carbon using the standard oxygen and carbon nuclear reactions [$O^{16}(d,p_0)O^{17}$], [$O^{16}(d,p_1)O^{17}$] and [$C^{12}(d,p)C^{13}$]. The spectra were fitted with the SIMRA software.

3. Results and discussion

3.1 Ablation process/FLG structure

For the present purpose the discs of two kinds of graphite have been used for the mechanical ablation: Highly oriented pyrolytic graphite (HOPG) and Thermal pyrolytic graphite (TPG).

According to the previous general procedure of mechanical exfoliation the graphite disc is ablated on a rough glass surface using an automatic set-up, while the ablated (exfoliated) product is detached from the surface using a stream of solvent (ethanol or possibly water) [1]. The ablation set-up is however modified compared to the one used for pencil lead (see *Methods*). Figure 1 is a draft representing the ablations of both: pencil lead and graphite disc. The presence of an inorganic binder such as kaolin in pencil lead is not the only difference between the lead and graphite disc. It has been found that the final products, FLG flakes, vary in term of morphology, more precisely by the size and flakes' edges. The FLG flakes from the exfoliation of lead demonstrate often "Multi-step" structure (fig. 1 a and 2a), which was previously confirmed by TEM microscopy, and through the stabilization of metal NPs [13], also in the self-assembling process where "3D-like particle" behavior was observed [14]. The multi-step structure of the edges exists but it is much less observed in the case of FLG originating from the graphite discs. We ascribe this difference to a different arrangement of the initial graphite species in the lead vs. graphite disc. The surface of ablation is parallel to the macroscopic surface of graphitic disc but is heavily parallel to the graphitic moieties in lead. In the latter, the graphitic species are relatively smaller and randomly arranged under a different angle as spaced by inorganic binder.

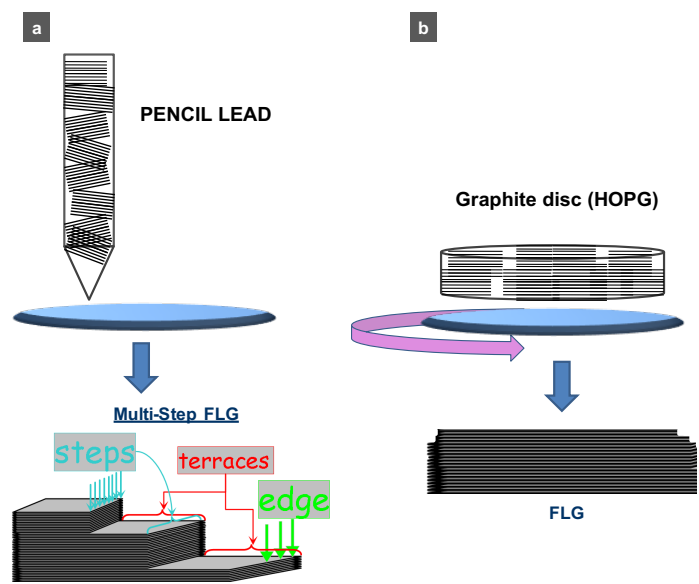

**Figure 1.** Draft of ablation process over a rough glass surface (**a**) of pencil lead and (**b**) of graphite disc; related FLG morphologies (lead-FLG with highly multi-step structure vs. graphite-FLG).

Figure 2 shows the related TEM micrographs of the edges of FLGs exfoliated from both precursors, where, additionally, the number of layers can be counted. In a similar way to the pencil lead ablation, the ablation of graphite was followed by a step of sedimentation, which aimed to separate the heavily exfoliated part of graphite. Likewise, the final supernatant (after few hours of sedimentation) contains FLG flakes with a number of sheets rarely exceeding 10, while the settled down part is multi-layer graphene (MLG) with a number of sheets up to 40. Unlike FLG from a pencil lead, no purification treatment is required in the case of graphite-FLG since no inorganic binder is present . The overall yield of graphite- FLG product reaches around 60-70%. The solvent, in this case ethanol, can be next removed for instance by standard rotary evaporator (figure 3a), without inducing excessive stacking of the flakes. We find it important to highlight this simple and fast step, since no literature reports such evaporation for graphene based materials. This kind of evaporation would seem to favor the π-π stacking between the FLG sheets, it can however be stopped before the solvent is totally removed leaving a thin layer of the solvent adsorbed on the FLG surface (due to the low pressure, the applied drying temperature can be highly decreased as well compared to standard drying). The remaining layer, in this case ethanol, helps afterword the dispersion of FLG in polar solvents as required for the introduction of FLG into polymers.

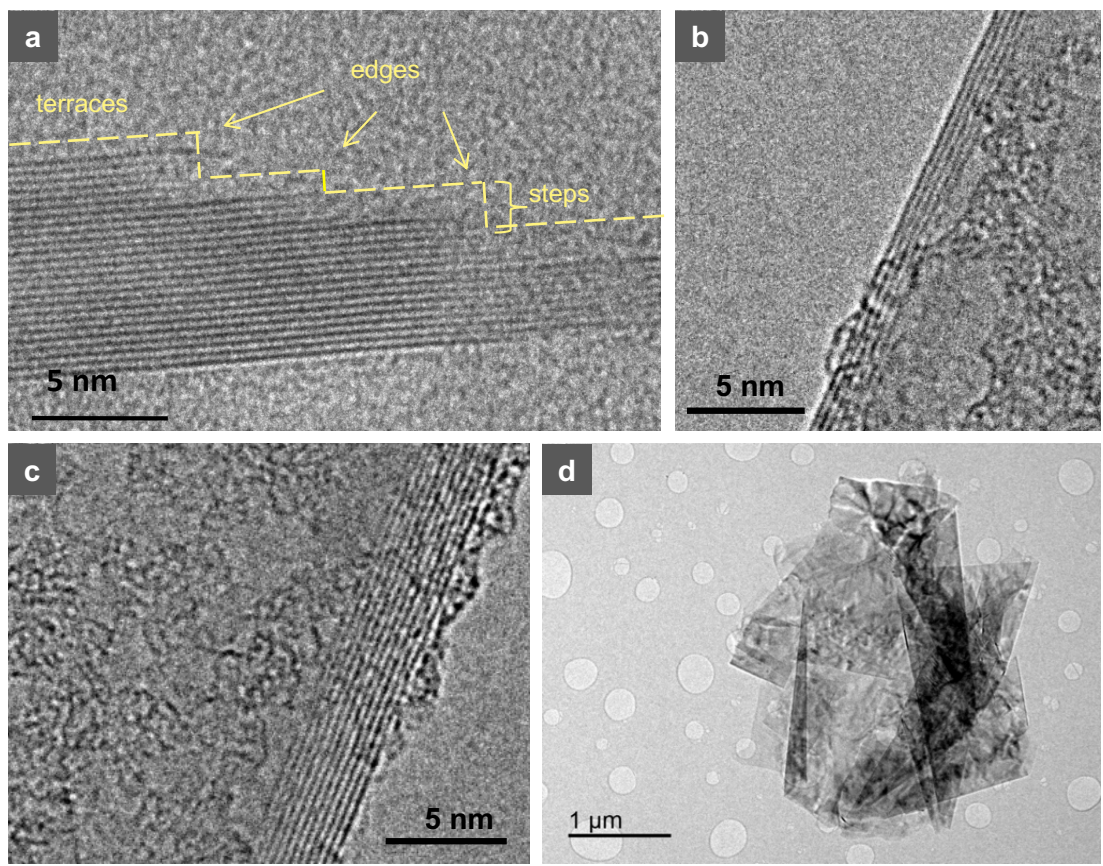

**Figure 2.** TEM micrographs of (**a**) thicker flake of lead-FLG with Multi-step structure, (**b** and **c**) variable number of sheets in graphite-FLG from the supernatant and settled-down part respectively, (**d**) overall view on the assembly of graphite- FLG flakes.

The high purity of graphite-FLG product was confirmed by XPS, Raman and TGA analysis (Figure 3 c-e). As it can be seen, the thin and relatively symmetric C1s peak at c.a. 284.5 eV in the XPS spectrum reveals a high content of sp$^2$ carbon and a very low content of sp$^3$ carbon and oxygen-bonded C whose signatures appear at higher binding energy, above 285 up to c.a. 290 eV. On the contrary, a clear π-π* transition loss peak at 291 eV revealing easy electron delocalization can be observed [15]. The low ratio of relative D and G peaks' intensity $I_D/I_G$ = 0.13 in Raman spectra confirms a very high content of undisturbed conjugated cyclic carbon. The shape of 2D peak in selected flakes at c.a. 2700 cm$^{-1}$ suggests few but no more than 5 layers in the analyzed flake (the intensity of the broad part of the peak at lower wavenumber is significant) [16]. Likewise, graphite-FLG oxidizes at relatively high temperature (under air), around 700°C, while, as expected, it starts to be oxidized 30-40 °C earlier the initial graphite-before the ablation.

3.2 FLG in nanocomposites

The rotary assisting-evaporated graphite-FLG was next used as an additive filler with 0.1 and 0.3 wt. % in the poly(trimethylene 2,5-furanoate), (PTF), where very high dispersion of FLG was observed after nanocomposite formation using an *in situ* polymerization process. The polymerization was preceded by dispersing the FLG in the monomer [17]. It was found that the addition of 0.3 % of FLG (PTF-FLG) does not increase the electrical conductivity, it slightly decreases oxygen transmission while thermal conductivity and elongation at break are significantly improved.

For a nanocomposite based on poly(trimethylene tetraphtalate) containing FLG from ablation of lead (PTT-FLG) different effects were observed. Here, for the same FLG content (0.3%) the elongation at break is strongly decreased ($\varepsilon_b$[%] is 2.28 vs. 178.32 for pure PTT), while on the contrary the permeability behavior is excellent and the $O_2$ and $CO_2$ transmission of PTT-FLG drop to negligible values from 74.3 to 2.3 and from 649.6 to 36.7 cm$^3$/m$^2$ × 24h, respectively.

Although the two polymers (PTF, PTT) are different and the interactions between them and FLG may vary, the signature of FLG morphology in the two composites is clear and suggests higher aspect ratio (lateral size vs. thickness) of FLG originating from pure graphite. The significant aspect ratio should indeed provide a higher elongation at break and possibly a creation of some preferential 2D paths of percolation in the graphite-FLG composite. Locally, the PTT is less charged with FLG, which permits however gas passing (nanocomposite films of 5 cm$^2$ were subjected to permeability measurements). On the contrary, the lower size and edges rich lead-FLG is highly and more homogenously distributed in the nanocomposite and its multi-step structures with inequivalent edges (and so inequivalent graphitic planes) helps to inhibit the permeability of gases at higher overall volume (surface and thickness), similar to a cascade-like barrier.

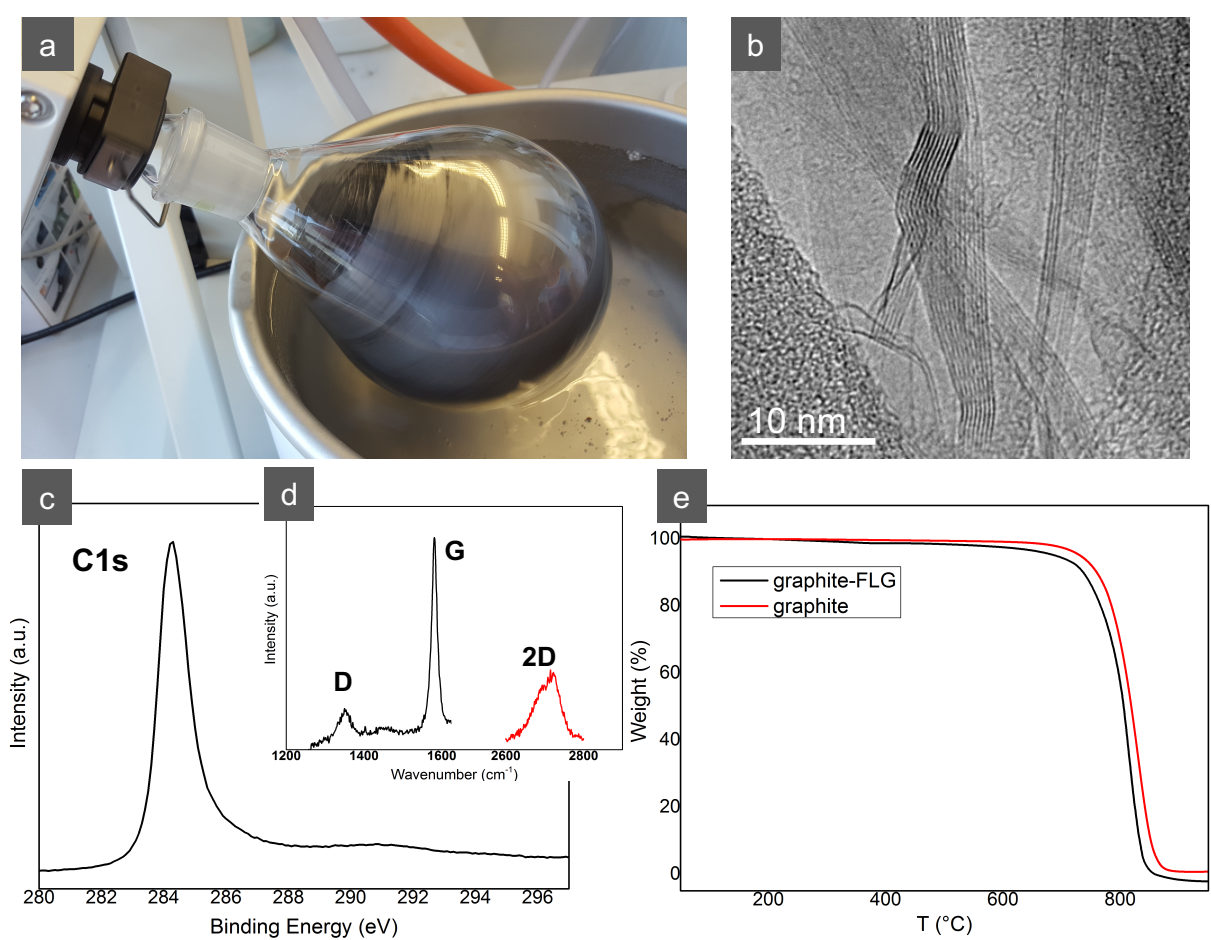

**Figure 3.** (**a**) removal of the solvent from FLG suspension by rotary evaporator, (**b**) TEM micrograph of graphite-FLG embedded into polymer (PTF) showing different number of edges/sheets in particular FLG flakes (3-9 sheets), (**c**) C1s XPS spectrum of graphite-FLG, (**d**) Raman spectrum of graphite-FLG, (**e**) TGA curves of graphite and graphite-FLG.

3.3 FLG conductive layer/film/electrode

The higher lateral size of the graphite-FLG flakes and structure homogeneity could be observed by SEM microscopy (figure 4 a,b) and also by investigation of the transparency-conductivity trade-off in FLGs films (figure 4 d-f) [18]. The FLG films/layers were formed by deposition of both FLGs over transparent substrates via hot spray technique, followed by annealing treatment in Ar. The annealing process allowed to remove ethanol and other possible impurities adsorbed on the FLGs flakes surface, which resulted in a decrease of the resistance. Both FLGs were deposited with minimum but high enough quantity to obtain the conductive layer. The layer formed from graphite-FLG shows a transparency up to 74% for a same range of conductivity (Rs ≈ 21 kΩ/sq) than FLG from the lead (Rs ≈ 15 kΩ/sq), while the transparency of the latter does not exceed 35%. It is worthy to note that conductivity measurements have been done by the four points probe method (FPPs). The Hall Effect method measurements performed previously (under $N_2$) for lead-FLG showed lower Rs by one order of magnitude, 1 kΩ/sq [19]. These results show that charge transport properties are more hindered in the case of lead-FLG compared to graphite-FLG: this is first due to the lower aspect ratio and, secondly, to the multi-step structure of the former. The conductivity of the films were calculated to be σ = 3.7 S/cm and 5.1 S/cm for graphite-FLG and lead-FLG respectively, while considering the transparency-conductivity figure of merit, the calculated "conductivity of transparency" values are $σ_{gt}$ = 61 S/cm and 41 S/cm for graphite-FLG and lead-FLG respectively [18].

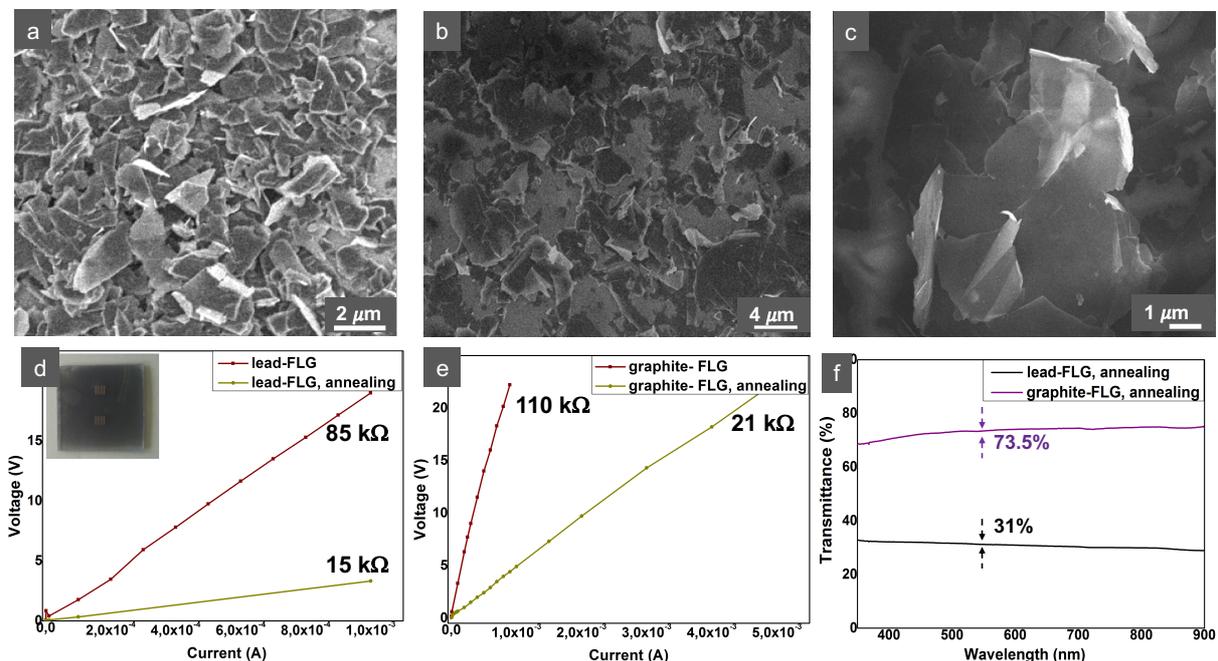

**Figure 4**. (**a-c**) SEM micrographs of sprayed layers: (**a**) lead-FLG, (**b, c**) graphite-FLG; (**d, e**) the I(V) curves obtained by FPPs method for lead and graphite-FLGs layers (without and after annealing), respectively, (**f**) transmittance curves of lead and graphite FLGs layers, for comparable conductivities of the films (Rs = 15kΩ/sq and 21 kΩ/sq, respectively).

The higher aspect ratio and, more precisely, the high lateral size of graphite-FLG flakes allows percolation and transport through the macroscopic surface (in 2D) at lower surface coverage. This entails the higher overall transparency.

In order to get more information about the morphology of the graphite-FLG layer prepared by hot-spray, the thickness of the film was measured by profilometry and the volume density (carbon) was determined afterwards by Nuclear Reaction Analysis (NRA). We find interesting to highlight at this point the use of NRA, since NRA is rarely applied for analysis of such relatively thick films prepared by high-scale method such as spraying. Figure 5a displays the NRA spectra obtained for conductive graphite-FLG film with $^{12}C (d, p)^{13}$ signal at c.a. 2735 keV [20].

Due to the resolution of NRA analysis (up to few μm), the depth profile includes the FLG film and $SiO_2$ substrate displaying two oxygen peaks for the two reactions. It was calculated from SIMRA software that the cross-sectional surface density (A) contains *6.25 × 10$^{17}$ atoms/cm$^2$*, while the average thickness of the film ($T_f$) is *130 nm* according to the profilometer measurements. Based on these two values, the volume density of the FLG film ($\rho_f$) can be calculated according to equation *1*:

$$\rho_f = A * M/(N_A * T_f) = 6.25 \times 10^{17} * 12/ (130 \times 10^{-7} * 6.023 \times 10^{23}) = 0.96 \text{ g/cm}^3 \quad (1)$$

where: $N_A$ is Avogadro number (*$6.023 \times 10^{23}$ atoms/mol*), M is the molecular weight of carbon (*12 g/mol*).

The obtained volume density of FLG film, $\rho_f$, is much lower than the volume density of graphite (~ *2.25 g/cm³*) indicating that more then half of the volume of the FLG films are voids (~ 57%).

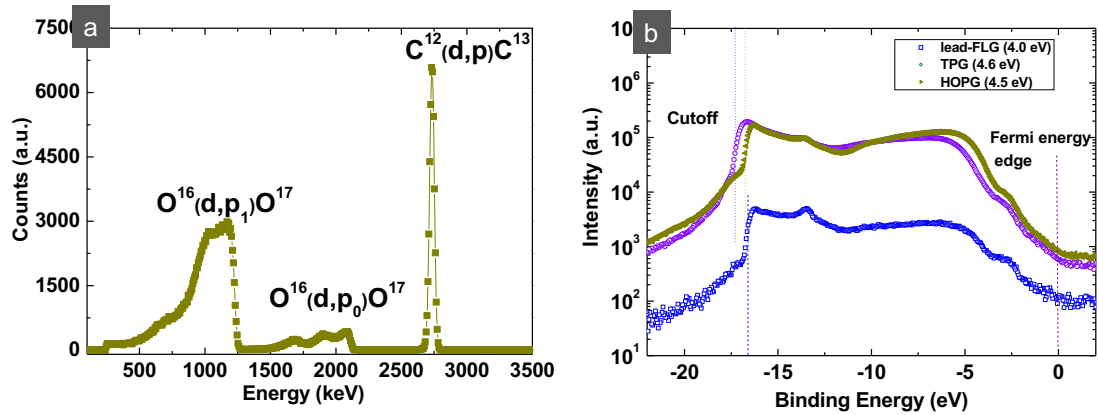

**Figure 5**. (**a**) NRA spectra obtained for graphite-FLG film (electrode) with carbon $^{12}$C at 2770 keV, (**b**) UPS work functions of lead- and graphite-FLG samples.

Despite a significant amount of voids, the FLG film is conductive meaning that the amount and arrangement of the FLG flakes are sufficient to form conductive paths. On the other hand, regarding the average thickness of the film and amount of the voids it is important to note that the overall transparency of the film is also due to the voids and not only to the low thickness of the FLG flakes themselves. The transmittance values obtained by UV-Vis spectroscopy are measured for micrometric spots that cover relatively high surface including the FLG flakes and the voids. The low homogeneity and still relatively low transparency of such prepared FLG film make it insufficient for applications like transparent conductive films/electrodes (TCFs). However, conductive layers/film with other useful properties such as hydrophobicity, anti-static, thermal conductivity etc… can be easily obtained. Yet, further optimization of the FLG-graphite structure by additional separation methods and appropriate deposition would lead to better transparency-conductivity- homogeneity characteristics of the films. Meantime, in view of the potential application of mechanically ablated FLGs as electrodes/conductive layers, the work functions (Φ) measurements were additionally performed by UPS method. The Φ values were determined by subtracting the difference between the UPS secondary cutoff (cutoff) and Fermi edges $E_f$. The obtained Φ for lead-FLG, 4.0 ± 0.1 eV, is much lower than the Φ of graphite/graphene, ~ 4.6 eV [21], which could be related to the purification step and the structure itself. The Φ of TPG-FLG and HOPG-FLG, 4.6 ± 0.1  and 4.5 ± 0.1  eV respectively, corresponds to Φ of graphite/graphene and are almost equivalent to Φ of ITO (4.7 eV) showing that this FLG could be envisaged for applications in optoelectronics if naturally the conductivity/transparency trade-off is resolved. On the contrary, the FLG-lead will be more appropriate as electrode with a low Φ, for instance as a solution-processable material that can replace metal-based counter electrodes, cathodes, in organic photovoltaic (OPV) cells (Ag, Al electrode: 4.0-4.3 eV). The metal electrodes are in general deposited by costly evaporation methods such as sputtering.

**4. Conclusions**

The significant modifications of morphology and some related properties of FLG obtained by mechanical ablation (exfoliation) of graphite are observed compared to the one obtained earlier by ablation of pencil lead. The new approach is a high yield and simple method with a simplified set-up and absence of additional purification steps. The new product (graphite-FLG) are flakes of few-several sheets with much higher aspect ratio and more homogenous structure compared to lead-FLG. Due to the different arrangement and purity of graphitic entities in the initial materials, larger FLG flakes with homogenous size of sheets and equivalent edges are obtained, while multi-step structure FLG flakes from pencil lead were often observed. The two different structures of graphite-FLG and lead-FLG reflect different signatures in nanocomposites (in polymers) and in conductive layers. The enhanced conductivity-transparency properties of film and increased elongation at break in nanocomposites of graphite-FLG compared to lead-FLG are measured. In turn, the increased number of effective edges in the multi-step FLG flakes (lead-FLG) is advantageous for the reduction of gases permeability in nanocomposites but also e.g. for the enhancement of dispersion and stabilization of metal nanoparticles for catalytic application. NRA analysis of the graphite-FLG films was used to get inside cross-sectional density and porosity of the film, and UPS analysis to check the work function values for both graphite- and lead-FLGs. The work function also varies for both FLGs placing them in different potential electrode applications. Concerning the lower aspect ratio and multi-step structure lead-FLG flakes, they exhibit less 2D and more 3D character, which can be highly suitable for the applications where the presence of active/reactive edges is beneficial, e.g. in catalysis [22] or supercapacitors' electrodes [23].

**Author Contributions:** A. A. Pirzado: preparation of FLG and FLG films, selected analysis, F. Le Normand: NRA analysis and related discussion, T. Romero: SEM microscopy, modification of ablation set-up, S. Paszkiewicz: preparation and analysis of nanocomposites, V. Papaefthimiou: UPS analysis, D. Ihiawakrim: TEM microscopy, I. Janowska: PI/coordinator of work, wrote the manuscript.

**Funding:** This research was funded by CONECTUS ALSACE (2010-2012) .

**Acknowledgments:** The *Conectus Alsace* is acknowledged for the financial support. Dr. Cuong Pham-Huu is acknowledged for help to get the Conectus support. Higher Education Commission, Pakistan is acknowledged for the financial support for A. A. Pirzado. Y. Le Gall and D. Muller (Icube/MaCEPV) are acknowledged for NRA analyses.

**Conflicts of Interest:** On behalf of all authors, the corresponding author states that there is no conflict of interest.